\documentclass[reprint,superscriptaddress,floatfix,aps,prb]{revtex4-2}
\usepackage{header}

\begin{document}

\title{Finite-temperature spin diffusion in the two-dimensional XY model}

\author{Erik Fitzner}
\affiliation{Institut f\"ur Theoretische Physik, Universit\"at T\"ubingen, Auf der Morgenstelle 14, 72076 T\"ubingen, Germany}
\author{Byungjin Lee}
\affiliation{Department of Physics, Korea Advanced Institute of Science and Technology, Daehak 291, Daejeon, 34141, Republic of Korea}
\author{Junhyeok Hur}
\thanks{Current address: Institut f\"ur Physik, Universit\"at T\"ubingen, Auf der Morgenstelle 14, 72076 T\"ubingen, Germany}
\affiliation{Department of Physics, Korea Advanced Institute of Science and Technology, Daehak 291, Daejeon, 34141, Republic of Korea}
\author{Minseok Kim}
\affiliation{Department of Physics, Korea Advanced Institute of Science and Technology, Daehak 291, Daejeon, 34141, Republic of Korea}
\author{Benedikt Schneider}
\affiliation{Arnold Sommerfeld Center for Theoretical Physics, 
Center for NanoScience,\looseness=-1\,  and 
Munich Center for \\ Quantum Science and Technology,\looseness=-2\, 
Ludwig-Maximilians-Universit\"at M\"unchen, 80333 Munich, Germany}
\author{Jae-yoon Choi}
\email{jae-yoon.choi@kaist.ac.kr}
\affiliation{Department of Physics, Korea Advanced Institute of Science and Technology, Daehak 291, Daejeon, 34141, Republic of Korea}
\author{Björn Sbierski}
\email{bjoern.sbierski@uni-tuebingen.de}
\affiliation{Institut f\"ur Theoretische Physik, Universit\"at T\"ubingen, Auf der Morgenstelle 14, 72076 T\"ubingen, Germany}


\begin{abstract}
    We present a combined theory-experiment study to quantify spin diffusion in the square lattice quantum spin-$1/2$ XY model at finite temperature. On the theory side, we leverage a recently developed dynamical high-temperature expansion method to faithfully capture the long spatiotemporal scales of the hydrodynamic regime. Experimental results are obtained from an optical lattice hard-core boson quantum simulator. The excellent agreement of spin diffusion constants marks a breakthrough in spin-transport beyond one dimension and for the quantitative validation of state-of-the-art quantum simulation platforms. We also provide theory predictions for future experiments on dynamic spin conductivity or anisotropy-induced integrability breaking.
\end{abstract}

\maketitle


\textit{Introduction.---}
Analog quantum simulation of many-body systems promises quantitative accuracy and the capability to obtain observables beyond the reach of numerical simulations on classical computers \cite{Altman2021, Bloch2012, Georgescu2014, TGrassRMP2025}. Fulfilling this promise requires careful navigation on the frontiers of both simulation approaches for cross-validation. One recurrent theme in this field considers the spatiotemporal diffusion process of a globally conserved quantity like particle number, charge or spin in a close-to-equilibrium situation. 
While the phenomenon of diffusion \cite{kadanoffStatisticalPhysicsStatics2000} admits an effective classical description and can be subsumed in an experimentally accessible diffusion constant, it emerges from the underlying quantum many-body dynamics. Its quantitative theoretical description in a chaotic many-body system therefore demands an accurate solution of quantum dynamics governed by the Schrödinger equation for large time and length scales \cite{bruusManyBodyQuantum2004} which represents a formidable challenge.

The above aspects render the problem of diffusion an excellent case for cross-validation between analog quantum simulation and numerical algorithm development. In the following, we will focus on the important case of spin diffusion. 
Although much experimental effort has been directed to one-dimensional (1D) spin systems and the question under which conditions diffusive spin-dynamics is realized at all ~\cite{Zotos1997,Znidaric2011,Steinigeweg2011,Bertini2021,Ronzheimer2013,Maeter2013,Xiao2015,Schneider2012,weiQuantumGasMicroscopy2022,Chen2026}, quantitative study of spin diffusion in 2D, which occurs much more generically, has received much less attention. One outstanding experimental study by Nichols et al.~\cite{Nichols2019} probed the equilibrium spin diffusion constant in a square-lattice Fermi-Hubbard model quantum simulator and mapped out its dependence on interaction strength. Remarkably, until now, seven years after the experiment, the obtained data could not be quantitatively reproduced theoretically~\cite{pham_spin_2026}. This highlights the shortcomings of state-of-the art numerical approaches to 2D quantum dynamics, and in particular the difficulty to account for the two vastly different spin transport time-scales in the Fermi-Hubbard model, namely spin-exchange and particle hopping. 
\begin{figure}[b!]
    \centering
    \includegraphics[width=86mm]{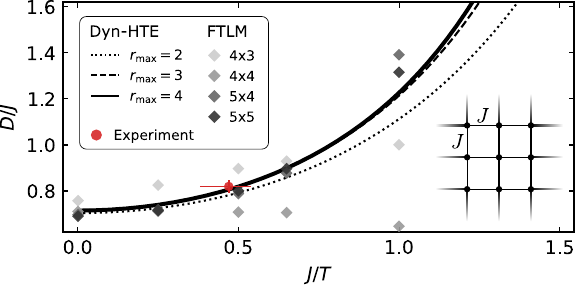}
    \caption{Temperature dependence of the spin diffusion constant $D$ for the square lattice XY model (inset). Theoretical results from Dyn-HTE (lines) are compared to FTLM (diamonds, on $L_x \!\times\! L_y$ lattice), the latter method is severely restricted by finite size effects for $T\lesssim J$ where Dyn-HTE converges well in the number of included frequency moments $r_{max}$. The experimentally determined $D$ (red dot, with errorbars) is significantly different from the theory value at $T=\infty$ but is in excellent agreement with the prediction at finite $J/T$ obtained by independent correlation-based thermometry.}
    \label{fig:D(T)}
\end{figure}

In this work we revisit the problem of equilibrium 2D spin diffusion theoretically and experimentally. For a fresh perspective, we focus on the simple and paradigmatic spin-$1/2$ XY model with nearest-neighbor exchange $J$ on the square lattice which we realize in the hard-core limit of a large-scale Bose-Hubbard quantum simulator. On the theory side, the degree of difficulty is controlled by the single free parameter in the problem, the dimensionless inverse temperature $J/T$. We demonstrate a novel theoretical approach to 2D spin diffusion based on the recently introduced dynamic high-temperature expansion (Dyn-HTE~\cite{Dyn-HTE_lett,Dyn-HTE_long}) which outperforms existing state-of-the-art techniques in terms of temperature range and accuracy. On the experimental side, we obtain the spin diffusion constant from the late-time decay dynamics of a domain wall state. By invoking independent correlation-based thermometry, we achieve excellent agreement between the measured spin diffusion constant and the theoretical prediction, see Fig.~\ref{fig:D(T)}. Our result thus presents an important milestone in the quantitative validation of analog 2D quantum simulation on the basis of quantum dynamics, highlights the importance of careful thermometry and enables various extensions to more complex observables or models. As an example for the latter, we theoretically study spin diffusion under the influence of controlled integrability breaking~\cite{Jung2006,jungSpinConductivityAlmost2007} through lattice anisotropy.

\textit{XY model and spin diffusion.---}
We consider a square lattice of localized spin-$1/2$ with unit spacing and nearest-neighbor coupling of in-plane spin components,
\begin{equation}\label{eq:H_XY}
    H = \sum_{\mathbf{r}} J_{x} S_{\mathbf{r}}^{+}S_{\mathbf{r}+\hat{\mathbf{x}}}^{-}+ J_{y} S_{\mathbf{r}}^{+}S_{\mathbf{r}+\hat{\mathbf{y}}}^{-} +\text{h.c.}\;.
\end{equation}
Here $S_\mathbf{r}^{\pm}=(S_\mathbf{r}^x \pm iS_\mathbf{r}^y)/\sqrt{2}$ with $S_\mathbf{r}^{x,y,z}=\sigma_\mathbf{r}^{x,y,z}/2$ spin operators at site $\mathbf{r}$ expressed via Pauli matrices $\sigma^{x,y,z}_\mathbf{r}$ and $J_{x,y}$ denote potentially anisotropic coupling strengths (we start with the isotropic case $J_x=J_y\equiv J$ and set $\hbar=1=k_B$). Unit vectors are denoted by $\hat{\mathbf{x}},\hat{\mathbf{y}}$. The model exhibits a $\mathrm{U}(1)$ spin rotation symmetry about the spin-$z$ axis so that the total magnetization $S^z=\sum_\mathbf{r}S_\mathbf{r}^z$ is conserved. The continuity equation on the lattice reads 
\begin{equation}\label{eq:ContinuityEq}
    \partial_tS^z_\mathbf{r} = -\!\nabla\cdot(j^{(x)}_\mathbf{r},j^{(y)}_\mathbf{r}) \equiv j^{(x)}_{\mathbf{r}-\hat{\mathbf{x}}}-j^{(x)}_{\mathbf{r}}+j^{(y)}_{\mathbf{r}-\hat{\mathbf{y}}}-j^{(y)}_{\mathbf{r}}\,,
\end{equation}
where spin(-$z$) currents follow from the equation of motion $\partial_tS^z_\mathbf{r}\!=\!i[H,S^z_\mathbf{r}]$, e.g.~the (bond) current between sites $\mathbf{r}$ and $\mathbf{r}+\hat{\mathbf{x}}$ reads
\begin{equation}\label{eq:Spincurrent}
    j^{(x)}_{\mathbf{r}}=-iJ_x (S^+_\mathbf{r}S^-_{\mathbf{r}+\hat{\mathbf{x}}}-S^-_\mathbf{r}S^+_{\mathbf{r}+\hat{\mathbf{x}}})\,.
\end{equation}
In other words, although spins are fixed in space, their mutual interaction leads to the spreading of $z$-magnetization throughout the system~\cite{Bertini2021,Steinigeweg2014,Bonca1995,Zotos1997}. In a (close-to) equilibrium state at temperature $T$, Fick's law governs the hydrodynamic relation between (weak) and spatially coarse-grained magnetization gradients and currents,
\begin{equation}\label{eq:Fick}
    \langle j^{(x)}_\mathbf{r}\rangle = -D \,\partial_x\langle S^z_\mathbf{r}\rangle\,,
\end{equation}
with $D$ the spin diffusion constant~\cite{Bertini2021,Bonca1995,Wienand2024,Nichols2019}. This equation forms the basis for the experimental determination of $D$ below. Here and in the following, we assume spin transport in $x$-direction and an infinite lattice. Note that $D$ does not depend on the sign of $J$ due to the bipartite nature of the lattice considered \footnote{This follows from a $\pi_z$ spin rotation on one sublattice which flips the sign of the Hamiltonian $H$ (interpreted as a sign-flip of $J$ from AFM to FM) but leaves the $zz$-DSF \eqref{eq:DSF} and thus, by Eqns.~\eqref{eq:Einstein_relation} and \eqref{eq:Cond_Spec}, also $D$ invariant.}.

\textit{Theoretical approach via Dyn-HTE.---}
From a theory perspective the spin diffusion constant is best calculated using the linear-response framework~\cite{bruusManyBodyQuantum2004} which requires to drive a spin current by a perturbation in the Hamiltonian and not by a magnetization gradient as in the experiment below. We thus perturb with a gradient of the Zeeman field $h^z_\mathbf{r}S^z_\mathbf{r}$ and use the Kubo formula~\cite{Bonca1995,Bertini2021}. The so-obtained dissipative part of the response function is known as spin conductivity, 
\begin{equation}\label{eq:Spincond}
    \sigma(\mathbf{k},\omega) \!=\! \frac{1\!-\!e^{-\omega/T}}{2\omega} \!\! \int_{-\infty}^{\infty}\!\!\! \mathrm{d}t\sum_{\mathbf{r}}e^{i\omega t-i\mathbf{k}\cdot\mathbf{r}}\langle j^{(x)}_\mathbf{r}\!(t)j^{(x)}_{\mathbf{0}}\rangle\,.
\end{equation}
This expression used the continuity equation \eqref{eq:ContinuityEq}, the fluctuation-dissipation theorem, and involves the time-evolved current operator $j^{(x)}_\mathbf{r}\!(t)$ in the Heisenberg picture~\cite{bruusManyBodyQuantum2004}. Its correlator is evaluated in thermal equilibrium without any gradients, $\langle\cdots\rangle=\mathrm{tr}[\cdots e^{-H/T}]/\mathrm{tr}\,e^{-H/T}$. Finally, $\sigma(\mathbf{k},\omega)$ relates to $D$ via the Einstein relation~\cite{Bertini2021},
\begin{equation}\label{eq:Einstein_relation}
    D \cdot \chi=\lim_{\omega\to0}\lim_{k\to0}\sigma(\mathbf{k},\omega)\,.
\end{equation}
Here and in the following $\mathbf{k}\equiv(k,0)$ is understood, $\chi$ is the static $zz$-spin susceptibility and the limits select the hydrodynamic regime. 

While Eqns.~\eqref{eq:Spincond} and~\eqref{eq:Einstein_relation} provide a formally exact route to calculate $D$, its practical evaluation is challenging, as the hydrodynamic limit requires access to the current correlator for sufficiently long time scales and system sizes. At infinite temperature, this problem is addressed by established approaches such as the recursion method~\cite{Tahir-Kehli1969,Morita1972,Wang2024} or, more recently, spin fRG~\cite{Kopietz2021}. At finite $T$, however, the situation is less well controlled. Here, the finite-$T$ Lanczos method (FTLM), based on quantum typicality~\cite{Hams2000,Reimann2018} and Lanczos recursion~\cite{Lanczos1950,Sandvik2010} was used ~\cite{Jaklic1994,Bonca1995,Shimokawa2025}. Our implementation of FTLM shows that its main restriction is the slow convergence of $D$ with system size $L_x \times L_y$ (restricted to $\sim\!25$ sites), see diamond markers in Fig.~\ref{fig:D(T)}. We conclude that FTLM is only suitable for high temperatures $T \gtrsim J$.
\begin{figure*}[t] 
    \centering
    \includegraphics[width=\linewidth]{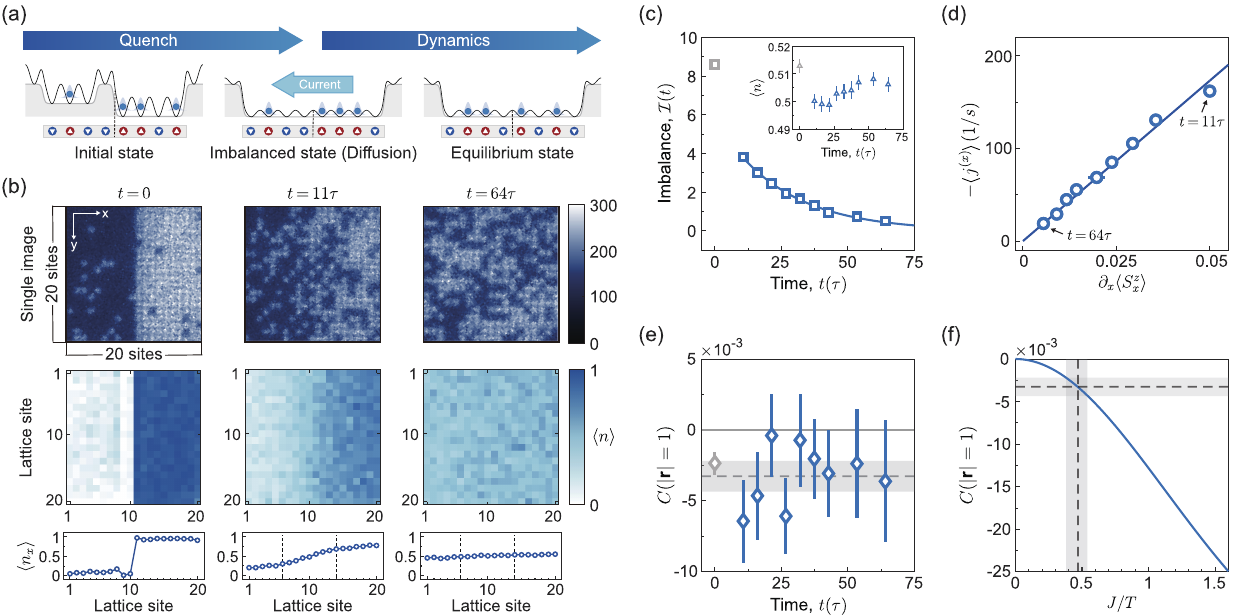}
    \caption{(a) Illustration of the spin diffusion experiment using hard-core bosons. A wall potential with height $44t_{\mathrm{BH}}$ is generated using a digital micromirror device (DMD). The total atom number is controlled by adjusting the offset potential in the left domain of the initial state through modification of the projected DMD pattern. (b) Time evolution of the system. The region of interest (ROI) for spin diffusion is $20\times20$. Raw fluorescence images with a photoncount color scale (top), averaged 2D densities (middle), and density profiles $\langle n_x\rangle$ averaged along the y-axis (bottom). The spin gradient is obtained from a linear fit (not shown) to $\langle n_x\rangle$ over the central 9 sites between the dashed lines. (c) Time evolution of the density imbalance. The solid line shows an exponential fit with no time or vertical offsets, from which the spin current $\langle j^{(x)}\rangle$ is obtained by differentiation. The initial time (gray point) is excluded from the fit. Error bars represent the standard error of the mean (SEM) and are smaller than the marker size. (Inset) Time evolution of atom density within the ROI. Atom density is conserved within $\langle n\rangle = [0.495, 0.515]$ during the dynamics. (d) Spin current as a function of the spin gradient $\partial_x\langle S^z_{x}\rangle$ at the center $x=10$. The solid line represents a linear fit used to extract the spin diffusion constant $D$. Vertical errorbars represent the propagated $1\sigma$ uncertainty of the derivative of the exponential fit in (c), while horizontal errorbars represent the $1\sigma$ uncertainty from the linear fit to the spin gradient. Each data point in (b)–(d) is obtained from $\sim 70$ realizations. 
    (e) Time evolution of the nearest-neighbor density correlator $C(\mathbf{r})=\langle(n_\mathbf{r}-\langle n_\mathbf{r}\rangle)(n_\mathbf{0}-\langle n_\mathbf{0}\rangle)\rangle$ with $|\mathbf{r}|=1$ averaged over the central $10\times10$ sites. The same data as in (b)-(d) are used, with density post-selection $\langle n\rangle=[0.475,0.525]$ within the reduced ROI. The dashed line represents the mean correlation value during the dynamics, and the gray shaded region indicates the propagated SEM, $S=\sqrt{\sum_{i=1}^Ns_i^2}/N$ with $N=9$.
    About 30 realizations are used for each point after post-selection. (f) Thermometry from nearest-neighbor correlation. The solid line shows the Dyn-HTE theory prediction. The horizontal dashed line and shaded region indicate the mean correlation value and its SEM from (e), while the vertical dashed line and shaded region indicate the extracted $J/T$ and its uncertainty. 
    }
    \label{fig:experiment}
\end{figure*}

We therefore apply the recently developed Dyn-HTE method~\cite{Dyn-HTE_lett,Dyn-HTE_long}. It operates in the thermodynamic limit and computes high-order HTEs for frequency moments of the dynamical structure factor (DSF),
\begin{equation}
S(\mathbf{k},\omega) = \int_{-\infty}^{+\infty} \! 
\frac{\mathrm{d}t}{2\pi}
        \sum_\mathbf{r}  
    e^{i\omega t-i\mathbf{k}\cdot\mathbf{r}}
    \!
    \left\langle S_\mathbf{r}^{z}(t)S_\mathbf{0}^{z}\right\rangle. \label{eq:DSF}
\end{equation}
The DSF can then be faithfully reconstructed from its resummed frequency moments by standard methods~\cite{Dyn-HTE_lett,Viswanath1994}. This works down to moderately low temperatures $T\simeq J/4$. Here we have extended Dyn-HTE to the spin-$1/2$ XY model and to order 14 in $J/T$, see also Ref.~\onlinecite{Fitzner2026} for a thermometry application in a Kagome Rydberg array.

We use the Fourier-transformed continuity equation~\eqref{eq:ContinuityEq} to connect DSF and spin conductivity \cite{Bonca1995}, 
\begin{equation}\label{eq:Cond_Spec}
    \sigma(\mathbf{k},\omega)=\omega(1-e^{-\omega/T})\pi S(\mathbf{k},\omega)/k^2,
\end{equation}
and obtain the HTEs for the frequency moments of $\lim_{k\to 0}\sigma(\mathbf{k},\omega)\!\equiv\! \sigma(\omega)$ by taking advantage of Dyn-HTE's real-space formulation, see Appendix for details. Using similar resummation and reconstruction techniques as for the DSF~\cite{Dyn-HTE_lett}, we access the limit $\omega \rightarrow 0$ and obtain $D(T)$ from Eq.~\eqref{eq:Einstein_relation} where we use a standard HTE for $\chi$. Our results are shown in Fig.~\ref{fig:D(T)} and we explicitly demonstrate convergence with respect to the number of (even) frequency moments, $r_{max}=2,3,4$ (various line styles) down to $T/J \simeq 2/3$ thereby outperforming FTLM. 


\textit{Experiment.---}
Using a quantum simulator consisting of ultracold bosonic $^7\mathrm{Li}$ atoms in an  optical square lattice \cite{Kwon2022}, we simulate the 2D isotropic Bose--Hubbard model,
\begin{equation}
H_\mathrm{BH}
\!= \!- \!\! \sum_{\mathbf{r}} t_{\mathrm{BH}}
(
b^\dagger_\mathbf{r} b_{\mathbf{r}+\hat{\mathbf{x}}}
\!+\! b^\dagger_\mathbf{r} b_{\mathbf{r}+\hat{\mathbf{y}}}
\!+\! \text{h.c.}
)
\!+\! \frac{U}{2} \!\! \sum_\mathbf{r} \! n_\mathbf{r}(n_\mathbf{r}\!-\!1).
\label{eq:Bosonic-Hamiltonian}
\end{equation}
Here, \(b_{\mathbf{r}}\) \((b_{\mathbf{r}}^\dagger)\) is the bosonic annihilation (creation) operator at site \(\mathbf{r}\), $t_{\mathrm{BH}}$ is the nearest-neighbor tunneling amplitude, and \(U\) is the on-site interaction energy. In the hard-core regime, \(t_{\mathrm{BH}} \ll U\), \(H_\mathrm{BH}\) maps onto the spin-\(1/2\) XY model \eqref{eq:H_XY}, with the mapping~\cite{Ronzheimer2013}
\begin{equation}\label{eq:Spinmapping}
    b^\dagger_\mathbf{r} = S^x_\mathbf{r}+iS^y_\mathbf{r},\;\; n_\mathbf{r}= b^\dagger_\mathbf{r} b_\mathbf{r}=S_\mathbf{r}^z-1/2,\;t_{\mathrm{BH}}=-J/2.
\end{equation}
The experiment begins by preparing a domain-wall state on an optical lattice with lattice spacing $a=752\, \mathrm{nm}$ \cite{Kwon2022} as shown in Fig.~\ref{fig:experiment}(a). A digital micromirror device (DMD) allows to generate arbitrary potentials and to tune the atom number in the left domain to satisfy the condition $\langle n\rangle\sim0.5$ while avoiding doublons in the right domain. Starting from the initial state in Fig.~\ref{fig:experiment}(b) ($t=0$), we rapidly switch the DMD potential from domain-wall to a box ($22\times22$ sites) and reduce the lattice depth to initiate the dynamics. During the evolution, the Bose-Hubbard parameters (see Appendix) are $t_{\mathrm{BH}}=h\times338(9)\,\mathrm{Hz}$ and $U/t_{\mathrm{BH}} = 30.2(8)$, under which the atom number is conserved within a $20\times20$ region of interest (ROI), see inset of Fig.~\ref{fig:experiment}(c). The ROI is chosen to minimize boundary effects of the box potential arising from the DMD point spread function \cite{hurStabilityManybodyLocalization2025,hur2024measuring,Kwon2026}.

To determine the spin diffusion constant using Fick's law in Eq.~\eqref{eq:Fick}, we measure the spin/density-current and -gradient at the center ($x=10$). As shown in previous work \cite{Nichols2019}, the spin current can be obtained from the time derivative of the density imbalance $\mathcal{I}(t) = \left\langle\sum_{x>10} n_{x} (t) - \sum_{x\leq10}  n_{x} (t)\right\rangle$, see Eq.~\eqref{eq:Spinmapping}. Here, $n_{x} (t)$ denotes the density averaged over the vertical $y$ direction at time $t$. The tunneling time $\tau=\hbar/t_{\mathrm{BH}} =0.47(1) \,\mathrm{ms}$ is shorter than the lattice quench duration of about $2\tau$, which is chosen to suppress band excitation \cite{Gericke2007}. 
For times beyond $11\tau$, we find empirically that the time evolution of the imbalance in Fig.~\ref{fig:experiment}(c) fits well to an exponential function with no offset. The current at the center of the box is obtained from the time derivative of the fitted imbalance curve as $-\langle j^{(x)}\rangle=-(1/2)(d\mathcal{I}/dt)$. The spin/density gradient $\partial_x\langle S^z_x\rangle=\partial_x\langle n_x\rangle$ is extracted from a linear fit to $\langle n_x\rangle$ over the central 9 sites, with the fitting range indicated in the bottom panels of Fig.~\ref{fig:experiment}(b). The measured spin current $-\langle j^{(x)}\rangle$ and the spin gradient $\partial_x\langle S^z_{x}\rangle$ fulfill a linear relation as predicted by Fick's law \eqref{eq:Fick}, see Fig.~\ref{fig:experiment}(d). A linear fit yields the diffusion constant $D=0.82(3)J$, the uncertainty denotes the $1\sigma$ fitting error and we set $\hbar=1$.

To determine the temperature of the close-to equilibrium state realized during the late-time domain wall melting dynamics, we measure the nearest-neighbor equal-time density correlator $C(\mathbf{r})=\langle(n_\mathbf{r}-\langle n_\mathbf{r}\rangle)(n_\mathbf{0}-\langle n_\mathbf{0}\rangle)\rangle$ \cite{Boll2016} with $|\mathbf{r}|=1$, as shown in Fig.~\ref{fig:experiment}(e). We average the correlator over the same time interval as used for the diffusion measurement and compare to the corresponding (Dyn-)HTE prediction of $\langle S_\mathbf{r}^zS_\mathbf{0}^z\rangle$ shown in Fig.~\ref{fig:experiment}(f) \footnote{Equal time correlators $\langle S_\mathbf{r}^zS_\mathbf{0}^z\rangle$ can be simply obtained from frequency sums of the Dyn-HTE for Matsubara correlators \cite{Dyn-HTE_long,Fitzner2026}}. We obtain $J/T=0.47^{+0.07}_{-0.09}$ and conclude that the experimental data point for $D(T)$ shown in Fig.~\ref{fig:D(T)} is in excellent agreement to the theory prediction.

\begin{figure}
    \centering
    \includegraphics[width=86mm]{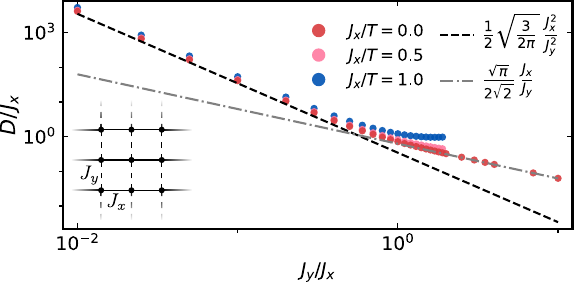}
    \caption{Scaling of spin diffusion constant $D(T)$ with lattice anisotropy $J_y/J_x$ (inset). In the limit $J_y/J_x \ll 1$, the diffusion constant follows a $D/J_x\propto(J_y/J_x)^{-2}$ scaling for all temperatures investigated. For $J_y/J_x \gtrsim 1$ and $J_x/T=0$, a crossover to $D\propto(J_y/J_x)^{-1}$ is observed. Dashed/dash-dottes lines indicate analytical estimates based on Gaussian approximations, see Appendix for details.}
    \label{fig:DS_scaling_Jy_Jx}
\end{figure}

\textit{Integrability breaking by anisotropy.---}
We finally generalize our theoretical considerations to the anisotropic case, i.e.~Hamiltonian \eqref{eq:H_XY} with $J_x\neq J_y$. We keep $x$ as the transport direction. Such anisotropy can be straightforwardly realized in optical lattice quantum simulators by independently tuning the laser intensities that define the optical lattice~\cite{hurStabilityManybodyLocalization2025}. In the limit $J_y/J_x\to 0$, the system reduces to decoupled chains, which are integrable and show anomalous non-diffusive transport $\sigma(\omega)\propto\delta(\omega)$~\cite{Cazalilla2011}. By continuously tuning the ratio $J_y/J_x$ we can therefore perform a controlled integrability breaking and monitor the crossover in spin transport.
Our Dyn-HTE results for the diffusion constant are shown in Fig.~\ref{fig:DS_scaling_Jy_Jx} for various $J_x/T$, where we find $D/J_x\propto(J_y/J_x)^{-2}$ in the limit of weakly coupled chains $J_y/J_x \ll 1$, in agreement with Fermi’s golden rule arguments~\cite{Jung2006,jungSpinConductivityAlmost2007}.
For $J_y/J_x\gtrsim 1$ and $T\rightarrow \infty$ we find a crossover to $D/J_x\propto(J_y/J_x)^{-1}$. Similar scalings have been reported for a ladder geometry~\cite{Steinigeweg2014}. For $D(T\!\rightarrow \!\infty)$ we derive simple but surprisingly accurate analytical expressions based on two complementary Gaussian approximations (dashed/dash-dotted lines and formulae in legend), see Appendix for details.

Note that similar integrability breaking in an XY-chain was recently studied experimentally~\cite{Chen2026} using Rydberg atoms with dipolar interaction. However, in this setup, the ratio of nearest- and integrability-breaking next-nearest-neighbor interaction ($J_2/J_1=1/8$) cannot be tuned. While late-time diffusive dynamics following a domain-wall quench was clearly observed, no quantitative determination of the diffusion constant was attempted.

\textit{Conclusion and outlook.---}
Interacting lattice spins are among the most simple and paradigmatic quantum many-body systems. Hence, diffusion of conserved magnetization is an elementary quantum-dynamic phenomenon in close-to equilibrium hydrodynamics. Despite this conceptual simplicity, to our knowledge  
no quantitative agreement between experimentally measured spin diffusion constants $D$ and theoretical predictions have been achieved so far beyond 1D. Historically, in the field of solid-state quantum magnetism, this relates to challenges in driving and measuring spin currents over sufficient time and length scales. One rare example for such an experiment is Ref.~\onlinecite{labrujereSpinDynamics3D1982}, but the obtained value for $D$ has still not been understood theoretically.

With the advancement of quantum simulation of spin models, the study of spin diffusion is rising to prominence again. Our work reports on the first match of a 2D spin diffusion constant between experiment and theory. This establishes a novel way for validation of quantum simulators. Moreover, the small error on the experimental value for $D$ renders it incompatible with the theoretical estimate at $T=\infty$ (c.f.~Fig.~\ref{fig:D(T)}) which highlights the need for careful thermometry and faithful inclusion of temperature on the theory side. This requires to go beyond simple methods that can only handle the large-$T$ regime. We have tackled this challenge by demonstrating that the recently developed Dyn-HTE method provides a powerful and quantitatively reliable approach to equilibrium spin diffusion down to moderately low $T/J$. For other applications, note that Dyn-HTE is readily applicable for arbitrary lattice geometries and also in 3D. 

Future work on the theory side could generalize Dyn-HTE to the Fermi-Hubbard model and finally provide a quantitative understanding of the spin diffusion measurement by Nichols et al.~\cite{Nichols2019}. On the experimental side, numerous opportunities await, even for the XY model. First of all, the $T$-dependence of $D$ could be resolved over a finite temperature range by controlled variation of entropy in the initial state. Second, using an anisotropic optical lattice, quantitative predictions on integrability breaking from Fig.~\ref{fig:DS_scaling_Jy_Jx} (and crossover time-scales to diffusive transport) could be investigated experimentally.  Another worthwhile experiment could measure frequency (or momentum) dependent spin conductivity \cite{andersonConductivitySpectrumUltracold2019}, for Dyn-HTE predictions on $\sigma(\omega)$ see Appendix. Finally, it would be interesting to leverage the flexibility of Rydberg-atom tweezer arrays \cite{chenContinuousSymmetryBreaking2023,Chen2026,TGrassRMP2025} to quantitatively study spin diffusion in other lattice geometries and under the influence of the native long-range (dipolar XY) exchange interactions if limitations on experimental time-scales permit.


\textit{Data availability.---}
The data and code that support the findings of
this work are openly available \cite{SpinDiff_Zenodo}.


\textit{Acknowledgements.---}
We thank 
Peter Kopietz and Christian Groß for useful discussions.
We acknowledge computational support by the state of Baden-Württemberg through bwHPC and the German Research Foundation (DFG) through Grant No.~INST 40/575-1 FUGG (JUSTUS 2 cluster).
This research is part of the Munich Quantum Valley Initiative, supported by the Bavarian Ministry for Arts and Science through the Hightech Agenda Bayern Plus.
We acknowledge funding from the Deutsche
Forschungsgemeinschaft (DFG, German Research Foundation) through the Research Unit FOR 5413 (project N2), Grant No.~465199066, and 
National Research Foundation of Korea (NRF) Grant under Project No. RS-2023-NR119928, RS-2023-00256050, RS-2025-02220735, and RS-2026-25469704.

\appendix
\section*{Appendix}

\begin{figure}
    \centering
    \includegraphics[width=86mm]{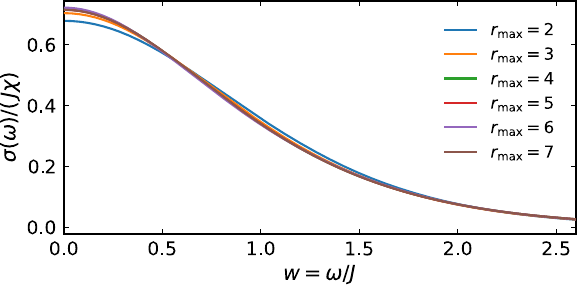}
    \caption{Frequeny-resolved spin conductivity $\sigma(\omega)$ at $T=\infty$ for the isotropic square lattice XY model, computed using an increasing number of moments $r_{max}$. The spin diffusion constant $D=\lim_{w\to0}\sigma(\omega)/\chi$ is well-converged in $r_{max}$.}
    \label{fig:Convergence_infiniteT}
\end{figure}

\textit{Spin conductivity from Dyn-HTE.---}
As shown in Eq.~\eqref{eq:Cond_Spec}, the spin conductivity $\sigma(\mathbf{k},\omega)$ is directly related to the DSF $S\left(\mathbf{k},\omega\right)$ which was the initial objective of Dyn-HTE \cite{Dyn-HTE_lett,Dyn-HTE_long}. For technical reasons, the method computes the HTE of the even frequency moments $m_{\mathbf{k},2r}$ ($r=0,1,2,...$) of the (even) relaxation function, $R_{\mathbf k}(w)=T A_{\mathbf k}(\omega)/(2\pi w)$, where $w=\omega/J_x$ and $A_{\mathbf{k}}(\omega)=(1-e^{-\omega/T})2\pi S\left(\mathbf{k},\omega\right)$ is the spectral function. Hence, the moments of $\lim_{k\to 0}\sigma(\mathbf{k},\omega)\!\equiv\! \sigma(\omega)$ are
\begin{equation}
    \langle w^{2(r-1)}\rangle \equiv \int_{-\infty}^\infty \!\!\!\! \mathrm{d}w\, w^{2(r-1)}\sigma(\omega)
    = \pi \frac{J_x}{T} \lim_{k\to 0} \! \frac{m_{\mathbf{k},2r}}{k^2}.
\end{equation}
In practice, the flexibility of Dyn-HTE rests on a real-space graph-embedding strategy of results pre-calculated for all possible lattice snippets (graphs) \cite{Dyn-HTE_long,Oitmaa2006}. This means that the moments are naturally available in real-space, $m_{\mathbf{r}\mathbf{r}^\prime,2r}$.  For translation invariant and inversion-symmetric systems, we set $\mathbf{r}^\prime=\mathbf{0}$ as the reference site and the $k\to 0$ limit can be evaluated explicitly as
\begin{align}\label{eq:k_limit}
    \lim_{k\to 0}\frac{m_{\mathbf{k},2r}}{k^2}&=\lim_{k\to 0}\frac{1}{k^2}\sum_\mathbf{r}\cos(\mathbf{k}\cdot\mathbf{r})\,m_{\mathbf{r}\mathbf{0},2r}\nonumber\\
    &\overset{r>0}{=}-\frac{1}{2}\sum_\mathbf{r}(r^x)^2m_{\mathbf{r}\mathbf{0},2r}\,,
\end{align}
where $\mathbf{r}=(r^x,r^y)$, $\mathbf{k}=(k,0)$ and we used $\sum_\mathbf{r} m_{\mathbf{r}\mathbf{0},2r}=0$ for $r>0$ from the static nature of the uniform susceptibility. Following the standard Dyn-HTE analytic continuation via a continued-fraction representation~\cite{Dyn-HTE_lett}, we obtain an estimate for $\sigma(\omega)$. Combined with Eq.~\eqref{eq:Einstein_relation}, this yields the spin diffusion constant
$
    D
    = \lim_{\omega\to 0}\sigma(\omega)/ \chi,
$
where we again employ the Dyn-HTE framework to obtain a numerical value for the static uniform ($zz$-)spin susceptibility, $
    \chi=\sum_\mathbf{r}\left\langle S_{\mathbf{0}}^{z}S^z_\mathbf{r}\right\rangle/T. $
In Fig.~\ref{fig:Convergence_infiniteT}, we show $\sigma(\omega)/(J\chi)$ at infinite temperature (where $1/\chi=4T$) for the isotropic square lattice, computed with increasing accuracy by including moments up to $r_{max}=2,...,7$. For $\omega\to0$, we obtain a well-converged value
\begin{equation}
    (D/J)^{\text{square lattice}}_{T=\infty} \approx 0.72\,,
\end{equation}
in excellent agreement with Ref.~\onlinecite{Morita1972}. For the two-leg ladder, we find
\begin{equation}
    (D/J)^{\text{ladder}}_{T=\infty} \approx 0.95\,,
\end{equation}
also in very good agreement with previously reported values~\cite{Steinigeweg2014,Wang2024}, supporting the reliability of our method.

Finally, note that HTE of moments $m_{\mathbf{r}\mathbf{r}^\prime,2r}$ are available up to order $(J_x/T)^{14-2r}$. For $T<\infty$, this means that the moment's expansion order decreases with increasing $r$ and we cannot faithfully resum the too short series for $r_{max}=5,6,7$. This explains the restriction to $r_{max}=2,3,4$ in Fig.~\ref{fig:D(T)}.
\begin{figure*}
    \centering
    \includegraphics[width=0.9\linewidth]{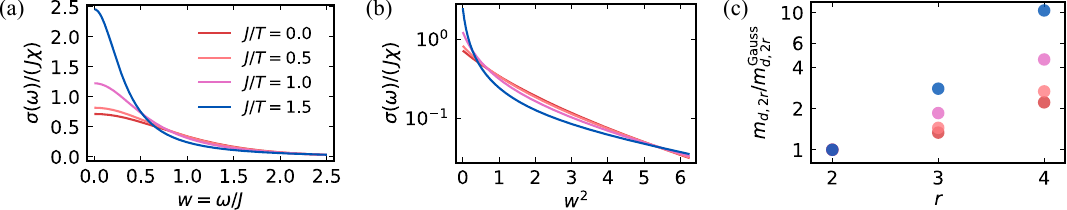}
    \caption{Deviations of $\sigma(\omega)$ from Gaussian behavior in the isotropic square lattice XY model. (a) Frequency dependence for different temperatures $J/T$. (b) The same data plotted as a function of $w^2$ on a logarithmic scale, where a Gaussian appears linear. (c) Ratio of higher-order frequency moments to their Gaussian counterparts.}
    \label{fig:Gauss_approx}
\end{figure*}

\textit{Gaussian approximation.---}
At infinite temperature and for simple models without multiple energy scales, it is often a good approximation to assume that $\sigma(\omega)$ is a Gaussian, see e.g.~Ref.~\onlinecite{Bonca1995} and Fig.~\ref{fig:Convergence_infiniteT},
\begin{equation}\label{eq:Gauss_approx}
    \sigma(\omega)=ae^{-b\omega^2}\,,
\end{equation}
where $a,b\geq 0$ are parameters left to be determined. Within this approximation, the only two independent frequency moments are $\langle \omega^0\rangle=a\sqrt{\pi/b}$ and $\langle\omega^2\rangle=\frac{a}{2b}\sqrt{\pi/b}$. The Einstein relation \eqref{eq:Einstein_relation} then yields 
\begin{equation}
    D_\text{Gauss}=\frac{a}{\chi}=\frac{1}{\chi\sqrt{2\pi}}\sqrt{\frac{\langle\omega^0\rangle^3}{\langle\omega^2\rangle}}\,.
\end{equation}
At infinite temperature, these frequency moments can be evaluated straightforwardly:
\begin{eqnarray*}
 \left\langle \omega^{2r}\right\rangle \! & \equiv & \int_{-\infty}^{\infty}\mathrm{d}\omega\,\omega^{2r}\sigma(\omega)\\
 & = & \frac{1}{2NT}\!\!\int_{-\infty}^{\infty}\!\!\!\mathrm{d}\omega\,\omega^{2r}\int_{-\infty}^{\infty}\!\!\!\mathrm{d}te^{i\omega t}\left\langle j^{(x)}(t)j^{(x)}(0)\right\rangle \\
 & = & \frac{1}{2NT}\!\!\int_{-\infty}^{\infty}\!\!\!\mathrm{d}\omega\int_{-\infty}^{\infty}\!\!\!\!\mathrm{d}t\left\{ \left(i\partial_{t}\right)^{2r} \! e^{i\omega t} \!\right\} \!\!\left\langle j^{(x)}(t)j^{(x)}(0)\right\rangle \\
 & = & \frac{\pi}{NT}\left\langle \left\{ \left(i\partial_{t}\right)^{2r}j^{(x)}(t)\right\} j^{(x)}(0)\right\rangle |_{t=0}\\
 & = & \frac{\pi}{NT}\left\langle [\dots[[j^{(x)},H],H]\dots]j^{(x)}\right\rangle,
\end{eqnarray*}
with a $2r$-fold commutator and  $j^{(x)}=\sum_\mathbf{r}j^{(x)}_\mathbf{r}$ the total (zero-momentum) spin current. This can be simplified as the product of two $r$-fold commutators~\cite{Viswanath1994},
\begin{equation}
\left\langle \omega^{2r}\right\rangle =\frac{\pi}{NT}(-1)^{r}\left\langle [\dots[j^{(x)},H],\dots]\cdot[\dots[j^{(x)},H],\dots]\right\rangle,
\end{equation}
and we insert $j^{(x)}$ above Eq.~\eqref{eq:Spincurrent} and $H$ from Eq.~\eqref{eq:H_XY} to find after lengthy but straightforward algebra
\begin{eqnarray}
\left\langle \omega^{0}\right\rangle  & = & \frac{\pi J_{x}^{2}}{8T},\label{eq:Moments_0}\\
\left\langle \omega^{2}\right\rangle  & = & \frac{\pi J_{x}^{2}J_{y}^{2}}{8T},\label{eq:Moments_2}\\
\left\langle \omega^{4}\right\rangle  & = & \frac{3\pi J_{x}^{4}J_{y}^{2}}{32T}+O(J_{x}^{3}J_{y}^{3})+O(J_{x}^{2}J_{y}^{4})\,.\label{eq:Moments_4}
\end{eqnarray}
In the following, we use these moments to rationalize the anisotropy scaling of the diffusion constant at infinite temperature within the Gaussian approximation (see Fig.~\ref{fig:DS_scaling_Jy_Jx}). We distinguish two cases, the close-to-integrable weakly coupled case $J_{y}\ll J_{x}$ and the opposite case $J_{y}\gtrsim J_{x}$ (which includes $J_{y}\gg J_{x}$). We start with the second case.

For $J_{y}\gtrsim J_{x}$ we expect ordinary diffusive transport and the Gaussian approximation \eqref{eq:Gauss_approx}
to be valid~\cite{Bonca1995}. It then follows, with the pre-calculated moments \eqref{eq:Moments_0}--\eqref{eq:Moments_2} and $1/\chi=4T$,
\begin{equation}
D=\frac{1}{\chi\sqrt{2\pi}}\sqrt{\frac{\left\langle \omega^{0}\right\rangle ^{3}}{\left\langle \omega^{2}\right\rangle }}=\frac{\sqrt{\pi}}{2\sqrt{2}}\cdot\frac{J_{x}^{2}}{J_{y}}\simeq0.63\cdot\frac{J_{x}^{2}}{J_{y}}\,,
\end{equation}
shown as dash-dotted line in Fig.~\ref{fig:DS_scaling_Jy_Jx} which matches Dyn-HTE at $J_x/T=0$ within a few percent and explains the observed scaling.

For $J_y \ll J_x$ we follow the general strategy outlined in Ref.~\onlinecite{Jung2006}. We first consider the decoupled limit $J_y=0$, where the XY chain is integrable and the total spin current is conserved, $[j^{(x)},H]=0$. As a consequence, the spin conductivity is purely ballistic,
\begin{eqnarray}
\sigma(\omega)\big|_{J_y=0}
&=& \frac{1}{2NT}\int_{-\infty}^{\infty}\!dt\,e^{i\omega t}\langle j^{(x)}(t)j^{(x)}(0)\rangle \nonumber \\
&=& \delta(\omega)\,\frac{\pi J_x^2}{8T}\,.
\label{eq:sigma_chain}
\end{eqnarray}
A weak transverse coupling $J_y \ll J_x$ breaks integrability and broadens the $\delta$-peak. We model this by a Lorentzian of width $\Gamma(\omega)\propto J_y^2$ representing the scattering rate,
\begin{equation}\label{eq:Lorentzian}
\sigma(\omega)=\frac{1}{T}\frac{\Gamma(\omega)}{\left(8\Gamma(\omega)/J_x^2\right)^2+\omega^2}\,,
\end{equation}
which reduces to Eq.~\eqref{eq:sigma_chain} in the limit $\Gamma\to 0$. The diffusion constant then follows as
\begin{equation}
D=\frac{\lim_{\omega\to0}\sigma(\omega)}{\chi}=\frac{ J_x^4}{64\,\chi\,\Gamma(0)T}\,.
\label{eq:Dxx_weak-Jy}
\end{equation}
For $\Gamma(\omega)\ll J_x^2$, the frequency moments of Eq.~\eqref{eq:Lorentzian} can be approximated to leading order in $J_y$ as
\begin{equation}
\langle \omega^{2r}\rangle \simeq \int_{-\infty}^{\infty} d\omega\, \omega^{2r-2}\, \Gamma(\omega)/T, \;\;(r=1,2,...).
\end{equation}
The moments \eqref{eq:Moments_2} and \eqref{eq:Moments_4} are sufficient to evaluate the parameters of a Gaussian ansatz for the scattering rate, $\Gamma(\omega)=a' e^{-b'\omega^2}$~\cite{Jung2006}. We obtain
\begin{equation}
\Gamma(0)=a'=\frac{T}{\sqrt{2\pi}}\sqrt{\frac{\langle \omega^{2}\rangle^{3}}{\langle \omega^{4}\rangle}}
= \frac{J_x J_y^2}{4}\sqrt{\frac{\pi}{6}}\,.
\end{equation}
Inserting this into Eq.~\eqref{eq:Dxx_weak-Jy}, we find
\begin{equation}
D = \frac{1}{2}\sqrt{\frac{3}{2\pi}}\,\frac{J_x^3}{J_y^2}\simeq0.35\cdot\frac{J_x^3}{J_y^2}\,,
\end{equation}
in good agreement with the Dyn-HTE results, see dashed line in Fig.~\ref{fig:DS_scaling_Jy_Jx}. 

The last example around Eq.~\eqref{eq:Lorentzian} implies that Dyn-HTE can access frequency dependence of $\sigma(\omega)$ beyond Gaussian lineshapes. In Fig.~\ref{fig:Gauss_approx} we investigate this for the isotropic case and analyze $\sigma(\omega)$ at finite $T$, see panel (a). Panel~(b) clearly reveals that deviations from a Gaussian shape are present even at $T=\infty$ and grow with $J/T$ while panel~(c) conveys the same message via the ratio of resummed frequency moments to their Gaussian counterparts.

\textit{Calibration of Bose-Hubbard parameters.---}
\begin{figure*}
\centerline{\includegraphics[width=\linewidth]{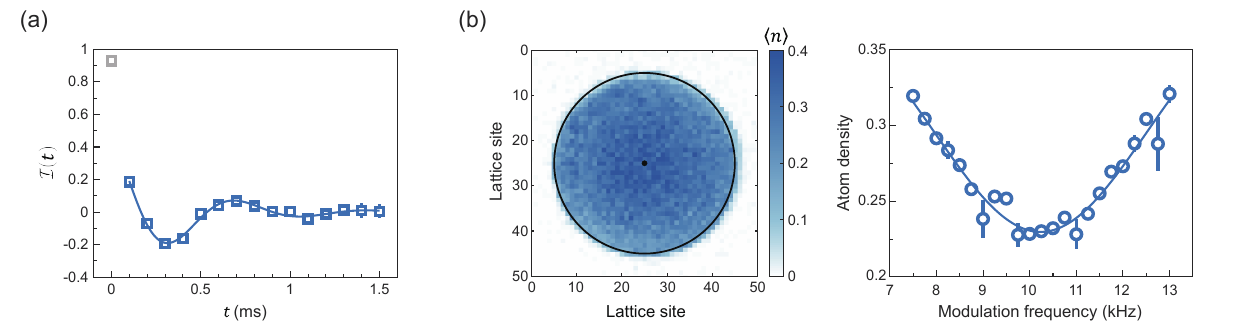}}
\caption{(a) Calibration of the nearest-neighbor tunneling strength $t_{\mathrm{BH}}$. The time evolution of the imbalance is fitted to a decaying Bessel function, $\mathcal{I}(t)=\mathcal{I}_0e^{-\gamma (t-t_0)}\mathcal{J}_0(4t_{\mathrm{BH}}(t-t_0))$, from which $t_{\mathrm{BH}}$ is extracted. Here, $\mathcal{I}_0$ is the imbalance amplitude, $\gamma$ is the decay constant, and $t_0$ is a time offset accounting for the dynamics during the quench. The solid line shows that the fit to $\mathcal{I}(t)$, excluding the  initial time (gray point). 8 realizations are used for each point. (b) Calibration of the on-site interaction $U$ using modulation spectroscopy. A ring potential is applied to confine the atoms within the region of interest (ROI). (Left) Density distribution averaged over all measurements; the circle indicates the ROI and the dot marks the center. (Right) Atom density within the ROI as a function of the lattice modulation frequency. The solid line shows a Gaussian fit, and 10 data are used for each point. Error bars in (a) and (b) represent the standard error of the mean (SEM).}\label{fig:ExpCalibration} 
\end{figure*}
To implement the mapping from the Bose-Hubbard Hamiltonian to spin-1/2 XY Hamiltonian, it is necessary to ensure that the system is in the hard-core regime ($t_{\mathrm{BH}}\ll U$). We set the lattice depth to $V_{lat}=10.8E_r$, where $E_r=h\times12.6\,\mathrm{kHz}$ is the recoil energy, in order to maximize the nearest-neighbor tunneling strength $t_{\mathrm{BH}}$ and minimize the effect of residual potential disorder ($\sim80\,\mathrm{Hz}$) in our system. We calibrate the tunneling strength using a stripe initial state by fitting the time evolution of the imbalance to a decaying Bessel function, $\mathcal{I}(t)=\mathcal{I}_0e^{-\gamma (t-t_0)}\mathcal{J}_0(4t_{\mathrm{BH}}(t-t_0))$ \cite{Wienand2024,hurStabilityManybodyLocalization2025}, as shown in Fig.~\ref{fig:ExpCalibration}(a). {The experimentally obtained value is $t_{\mathrm{BH}}=h\times 338(9)\, \mathrm{Hz}$, in good agreement with $t_{\mathrm{BH}}=h\times340\, \mathrm{Hz}$ from a numerical calculation based on Wannier functions~\cite{mitra2018quantum}. The initial point is excluded from the fit, since the shortest lattice quench duration without band excitation \cite{Gericke2007} is about $2\tau$, where $\tau=\hbar/t_{\mathrm{BH}} =0.47(1) \,\mathrm{ms}$ is the tunneling time. As a result, the dynamics during the quench cannot be neglected, and we include a time offset $t_0$ in the decaying Bessel fit.  \\
To realize the hard-core limit, we tune the magnetic field to achieve $t_{\mathrm{BH}}\ll U$ using a Feshbach resonance \cite{Chin2010,Amato-Grill2019}. The on-site interaction strength $U$ is measured via modulation spectroscopy \cite{Stoferle2004}. Owing to the short tunneling time, atoms escape from the region of interest during modulation. For this reason, a ring potential generated by a digital micromirror device (DMD) is used to confine the atoms, and the resulting density distribution is shown in Fig.~\ref{fig:ExpCalibration}(b). From the modulation spectroscopy, the on-site interaction is determined to be $U = h\times 10.2(0.02)\, \mathrm{kHz}$. The uncertainties of $t_{\mathrm{BH}}$ and $U$ represent the $1\sigma$ fitting error. These results confirm the system is in the hard-core regime, $U/t_{\mathrm{BH}} = 30.2(8)$.

\bibliography{references}

\end{document}